# Counting the homeless in Los Angeles County

**Richard Berk[1], Brian Kriegler[1] and Donald Ylvisaker[1]**

*University of California, Los Angeles*

**Abstract:** Over the past two decades, a variety of methods have been used to count the homeless in large metropolitan areas. In this paper, we report on an effort to count the homeless in Los Angeles County, one that employed the sampling of census tracts. A number of complications are discussed, includingÈ the need to impute homeless counts to areas of Èthe CountyÈ not sampled. We conclude that, despite their imperfections, estimated counts provided useful and credible information to the stakeholders involved.

## 1. Introduction

During the fall of 2004, the County of Los Angeles began a project to count the homeless. The effort was directed by the Los Angeles Homeless Services Authority (LAHSA), a City-County Joint Powers Authority independent of local government with a mandate to "address the problems of the homelessness on a regional basis" (www.lahsa.org). A non-profit firm, Applied Survey Research (www.appliedsurveyresearch.org), was awarded a contract to conduct the count, and brought us in as statistical consultants.

In this paper, we address the following questions. First, what was the political context in which the study was undertaken? Homeless counts are always controversial because of the range of stakeholders involved and the fiscal resources connected to the estimates. Second, what design was used to undertake the count? A number of approaches had been used in the past with widely varying success. Third, how well was the design implemented? Earlier studies made clear that there might be a lot of slippage between the research design and its implementation. Fourth, what were the findings? It was anticipated going in that Los Angeles County might well have the largest number of homeless individuals of any metropolitan area in the country. So, the stakes were high. And fifth, what more general lessons might be learned from the Los Angeles experience? There were numerous decision points during a lengthy study period that might now be reviewed with profit. We choose as well to take advantage of the uncommon opportunity to inspect the improvement in model prediction of homeless totals as the aggregation level increases, truth being known. Here, the size of the observed data set allows us to do this in a fairly substantial setting.

---

[1]University of California, Department of Statistics, Room 8125 Mathematical Sciences Building, Los Angeles, California 90095-1554, USA, e-mail: berk@stat.ucla.edu; bk@stat.ucla.edu; ndy@stat.ucla.edu

## 2. Past attempts to count the homeless

Estimates of the number of homeless depend on a clear definition of exactly what homelessness entails (Cordray and Pion [5]) and on effective methods to apply the definition. Both steps have been controversial (Burt [2], Chelimsky [4], Kondratas [7] and Rossi [11]). Neither is near resolution to our knowledge. Here we focus on methods to estimate the number of homeless because definitions of homelessness ultimately depend on what kinds of homelessness the political process is prepared to accept and pay for.

It is probably fair to say that efforts to apply rigorous methods for counting the homeless began with the study conducted by Peter Rossi and his colleagues for the city of Chicago in 1985 (Rossi et al. [10]). Prior to that time, homeless estimates were based on the judgement of homeless advocates or on interviews with local experts. In the Rossi study, Chicago was partitioned into city blocks and other geographical areas, together with sites like bus stations and airports, and characterized by the likelihood of finding homeless people in them. These areas were then sampled proportional to that probability. On a single night, enumerators were sent into the sampled areas to record all of the homeless individuals found. The street counts were supplemented with counts from homeless shelters.

A key concern with the Rossi approach is the ability of enumerators to obtain accurate information. Counting some people more than once is a problem in principle, but the dominant worry is failing to find homeless individuals who should be counted. In a series of studies conducted in collaboration with the Bureau of the Census during the night of March 20-21, 1990, a clever effort was undertaken to document the undercount (Taeuber and Siegel [13] and Martin et al. [9]). In each of five major cities, including Los Angeles,[2] approximately 60 "decoy" homeless individuals were put on the street in areas where other homeless were likely to be found. These confederates were instructed to report whether or not they had been counted by an enumerator. Between 22% and 67% of the decoys were counted, depending on the city (Wright and Devine [14]), and the likelihood of a significant undercount was confirmed.

Articles on the experience in each city are contained in a special issue of the *Evaluation Review*, edited by James D. Wright (1992). Each article addressed the reasons why enumerators might miss a large number of homeless individuals and what might be done to improve matters. A common thread through all of the articles was the need to better train and supervise enumerators. With better training and supervision, there was considerable optimism that the undercount could be substantially reduced.

The use of decoys also raised the possibility of adjusting the homeless counts. The proportion of decoys counted could be used in a capture-recapture sampling framework to "correct" a count (Laska and Meisner [8], Martin et al. [9] and Schindler et al. [12]). Thus one would infer that when the proportions found were smaller, the undercount was larger. However, some strong and untestable assumptions are required to produce reliable estimates.

There are alternatives to the Rossi strategy, but these too have limitations. One popular approach is to build on the capture-recapture idea. For example, Cowen [3] exploits information in the records of service-providers. If a person recorded to use a service on a given day is recorded to return the next day, that person has been "recaptured." So, the number "captured" on day 1, can be up-weighted by the inverse of the proportion "recaptured" on day 2 to arrive at an overall estimate. However, this would only apply to services that "empty out" between day 1 and day 2 (e.g., a soup kitchen), and assumes a) that homeless individuals do not enter or leave the catchment area between the two days and b) that all homeless individuals in the catchment area have the same non-zero probability of using the service in question.

---

[2] The others were New York, Chicago, Phoenix, and New Orleans.

Another popular strategy is to use statistical methods to impute what a set of homeless counts should be (Hudson [6]). For example, homeless counts at the census tract level can be regressed on other features of those census tracts. The resulting regression equation can be used to predict homeless counts when they are unavailable, as long as the same regressors can be utilized. Much the same approach can be used with data from service providers. The number of homeless individuals using the services during some specified periods and other variables (e.g., whether it is the holiday season) are taken as regressors when a model for the street counts is constructed. Future numbers of service users and other predictors can then be employed to impute street counts. Just as with capture-recapture methods, strong assumptions are required. For example, one must make the case that the regression model developed is in some sense "right." One must also assume that the initial regression results can be generalized to the census tracts or catchment areas for which counts need to be imputed.

In summary, all of the efforts to count the homeless are flawed in various ways. The methods apparently preferred by the Census Bureau depend on getting enumerators to do their job well (Martin et al. [9]). The alternatives depend on models of one form or another, and are only as good as the models themselves. The ways to make the census results more credible seem clear and, if resources are available, viable in practice. The modeling approaches are more difficult to improve because so many of the key assumptions are effectively untestable and/or lead to no practical remedies.

## 3. Setting the stage

Los Angeles County was established in 1850 as a relatively small agricultural and trading region along California's coast between Santa Barbara and San Diego. Over the next four decades, its boundaries changed several times until, in 1889, it arrived at its current dimensions. As of January, 2005, over 10 million people were estimated to live within Los Angeles County's 4,084 square miles, making the County the most populous in the nation with more residents than all but eight states.

Because of sheer size alone, the prospect of trying to count the number of homeless individuals was daunting. The usual sorts of political pressures complicated matters further. To begin with, the RFP from the county had been circulated in the early Spring of 2004, and several worthy proposals were submitted shortly thereafter. There was apparently some reluctance within LAHSA to proceed — the study would be costly for a county that was struggling to provide the range of public services that it had in the past. It was only after HUD insisted that future funding would depend on a credible measure of need that the contract was awarded. By then, several months were lost because the original HUD deadline remained.

Once an award was made, LAHSA mobilized quickly. Still, information necessary for an effective research design took several months to acquire. For example, a critical consideration in any counting effort would be information on where large numbers of homeless individuals were likely to be found, but moving beyond fragmented and anecdotal accounts took many weeks. Moreover, there were a number of important political considerations that were difficult to resolve. In particular, it was apparent from the start that it would be impossible with the resources available to send enumerators to all areas in the county. A sampling strategy was needed that was scientifically credible and yet managed to take the most important concerns of interested parties into account. With a large and heterogeneous set of stakeholders, arriving at a consensus set of concerns was never fully achieved.

These and other considerations suggested a design strategy that would balance several competing risks. For example, one could not take as the whole truth *a priori* accounts of where large numbers of homeless individuals would be found. At the same time, it would be foolhardy to ignore such information. In addition, prudence dictated building into the design as much flexibility as possible. Beyond a homeless estimate for the county as a whole, there was a stated need for counts from certain areas within the county. Yet little guidance was provided about what those areas were likely to be. One implication was that a way had to be found to provide counts in geographical areas to which enumerators were not sent.

## 4. Sampling design

The geographical area to be studied was determined by the boundaries of Los Angeles County, and the observational units were to be census tracts. Census tracts were selected because they are well defined and because they are a spatial unit for which considerable information was already available.

Los Angeles County is composed of 2,054 census tracts. Efficient use of resources dictated selecting a sample of tracts to produce an overall county estimate of the number of homeless. Homeless services for the County are in part organized into eight Service Provision Areas (SPAs). These contiguous areas, sporting such names as the San Fernando Valley, Metro Los Angeles, and the South Bay, suggest some of the unique features of the county and ensure that each has has strong advocates. Thus, an early attempt to sample in proportion to the size of homeless populations, as they were understood from earlier studies, foundered on politics, and the sample was stratified by SPA. Power calculations were undertaken to determine the approximate overall sample size required for desired precision.

There were discussions early in the process about providing estimated homeless counts for a variety of geographical areas within the county. The SPAs are an example, the incorporated cities in Los Angeles County are another and the five supervisorial districts yet another.[3] No decisions were made before the sample had to be drawn about specific areas within the county where estimated homeless counts would be needed.

Information provided by LAHSA indicated that 211 tracts could be anticipated to have a large number of homeless individuals and should be included with certainty among the tracts to be studied. From the remaining tracts, a stratified random sample of 299 was drawn. In this, each SPA formed a stratum with the number of tracts sampled in it proportional to the number of tracts it contained — SPAs with more tracts were more heavily sampled.

The design had positive features. First, by selecting with certainty the tracts anticipated to have large numbers of homeless individuals, one was assured of obtaining counts where they were most needed, at least to the extent that local knowledge of such locations was accurate (for information about this aspect, see Comment 1 of Section 11). Second, these tracts were then, in effect, removed from the population from which a random sample was to be drawn. This did, in fact, dramatically reduce the variability in counts in the population to be sampled, and led to a substantial gain in the precision of the results later obtained. Third, stratifying by SPA and sampling the number of tracts within each proportional to the actual number tracts in it did not differ markedly from a design which could be thought of as having been chosen proportional to size — politics did not interfere greatly with efficiency of estimation.

---

[3]In addition to Los Angeles city proper, Los Angeles County includes 87 incorporated cities such as Santa Monica, West Hollywood, Beverly Hills, Huntington Beach, Lynwood, Van Nuys, Burbank, and Inglewood. Los Angeles County is governed by a group of elected County Supervisors. Each represents an area of the County.

## 5. Implementation of the study

Field work was undertaken from midnight to the early morning hours on January 25, 26 and 27, 2005. Enumerators were paired with homeless individuals who were hired at $10 an hour to serve as "guides" in each census tract. Census tracts are large spatial areas, and it can be difficult to know where homeless individuals are likely to be found. The guides were to help in the search.

Shortly before the sample was drawn, we were told that the cities of Long Beach, Pasadena and Glendale would not cooperate in the study even though they were part of Los Angeles County and their numbers were to be included in the overall county estimate. Their counts would have to be imputed. This raised serious concerns because all three were alleged to contain substantial numbers of homeless, and we had no way of knowing how well counts obtained for the rest of the county could be extended into these areas.

After the enumeration was completed, we learned that 33 of the 39 tracts we had selected at random for enumeration in SPA 4 had been replaced by 33 others. The new 33 contained long sequences of consecutively numbered tracts, indicating proximity, and the decision was made to treat these 33 as if they had been selected with certainty. Thus, we considered that we had a random sample of size 6 from SPA 4 tracts, together with the original 53 tracts chosen with certainty, and now augmented by 33 others. Removing these tracts from the sample reduced the size of the estimation problem but also ensured that one had an inadequate sample size for dealing with it. SPA 4, Metro Los Angeles, is thought of as a high homeless area, and the segment of it attended to through sampling was not well served by the process.

We were not made aware of other major problems with implementation. Indeed, the procedure for obtaining sample counts was said to have proceeded smoothly.

## 6. Sampling estimates at the SPA level

Street counts were provided for each tract sampled or selected with certainty. Shelter counts then contributed to a total count. In this total, we considered that uncertainty attached only to the sampling process, and assigned no uncertainty to the sampled tract, selected tract and shelter counts obtained. The estimated number of homeless in each SPA attributable to its sampled tracts was computed as

$$(1) \qquad \hat{\tau} = \frac{N}{n} \times \sum_{i=1}^{n} c_i,$$

where $N$ is the number of sample-eligible tracts in a SPA, $n$ is the number of sampled tracts in which counts were obtained, and $c_i$ is the the number of homeless counted in sampled tract $i$. For example, if $N$ were 100, $n$ were 25, and the count from 25 sampled tracts was 200, the estimate for that SPA would be 800. Adding in the tract counts from the relevant excluded tracts and shelters produced an estimate for each SPA. Summing over SPAs produced an estimate for the county as a whole.

To obtain an estimate of the variance of the estimated count for a given SPA, first consider the (unknown) variance of a randomly selected tract count in the SPA using

$$(2) \qquad \mathrm{var}(c) = \frac{1}{n} \sum_{i=1}^{n} (c_i - \bar{c})^2,$$

where $\bar{c}$ is the mean of the counts for the SPA tracts and n is the total number of tracts. Then the (unknown) variance of the estimated count is

$$(3) \qquad \mathrm{var}(\hat{\tau}) = (N/n)^2 \times n \times \mathrm{var}(c) \times \frac{(N-n)}{(N-1)},$$

where $(N - n)/(N - 1)$ is the finite population correction. An estimate of this variance was obtained by replacing $\mathrm{var}(c)$ in (2) with its sample analogue. The square root of the estimate of $\mathrm{var}(\hat{\tau})$ was then the estimated standard error (SE) for each SPA's estimated count due to the sampling of tracts. The margin of error for each SPA was taken to be twice the estimated standard error for that SPA. For example, using the same numbers as above and supposing the sample estimate of $\mathrm{var}(c)$ to be 100, $\mathrm{SE}(\hat{\tau}) = \sqrt{16 \times 25 \times 100 \times .76} = 174$. If one further supposed the excluded tracts and shelters in the SPA contributed a count of 400 to go with the observed 800, the margin of error for that SPA would be expressed as $1200 \pm 348$.

The standard error for the county as a whole is the square root of the sum of the estimates of $\mathrm{var}(\hat{\tau})$ for each SPA. The margin of error for the county as a whole was taken to be twice the standard error for the county as a whole. This is usefully viewed as a percentage error compared with the estimated total.

## 7. Imputing homeless counts at the tract level

There was an interest in providing estimates of the number of homeless for all individual tracts in which counts were not available. Three strategies were considered for imputing the count in a given tract.

1. Using the SPA average—For each non-sampled tract, we use the mean count of the sampled tracts in the SPA in which the tract is found as the imputed value. This procedure appears throughout to serve as a benchmark for others. It is designated as Model 0.
2. Using matched tracts—For each tract, we match it to tracts with similar values for correlates of homelessness and use the average of the counts in the matched tracts as the imputed value. Use of the SPA average as above is just a special case of this — one views tracts as matched if they are in the same SPA.
3. Using correlates of homelessness—We construct a statistical model of how various features of tracts are related to homelessness and use the model's predictions as the imputed values. The previous strategy is a special case — categories of values being replaced by observed values.

When tract-level values are aggregated to large spatial units, all three methods, sensibly applied, will give similar results. The census tracts for which inputed values are needed are those that were not designated *a priori* as tracts with large number of homeless people. Insofar as this is correct, there is likely to be less variability in the counts in these tracts. After suitable aggregation, what variability there is tends to cancel out.

Such averaging out is not guaranteed to occur, however. It depends, in particular, on how many tracts are aggregated and which tracts they are. Should interest reside in a small number of tracts, the three methods can give somewhat different results (see Section 10 for some evidence on this point).

We rather reluctantly left Model 0 behind. The reasons for doing so included the following: (i) we had only a random sample of size six in SPA 4, and this was particularly unfortunate due to its inner city location; (ii) it was necessary to extrapolate to tracts in Long Beach, Pasadena and Glendale that were not in the sampling frame; and (iii) it was difficult to ignore possible homeless indicators that stakeholders "know" are of value in understanding homeless counts. Accordingly, the second and third strategies seemed likely to perform better than the first because they used more (presumably useful) information than the SPA membership noted in the first method. Between the matching and the modeling approaches, modeling seemed to perform a bit better in the manner in which available tract information was put to use.

Model 0 estimates of homeless by SPA are given in Table 1, together with their standard errors. Going forward, the latter figures provide the scales on which other estimates might be judged.

## 8. Results

### 8.1. *Estimates for the service provision areas and the county as a whole*

Table 1 gives Model 0 estimates aggregated to SPA and County levels. For specific tabled entries, recall the remarks about SPA 4 made in Section 5, and note Comment 3 of Section 11 as it pertains to SPA 2.

For the county as a whole, the figure of nearly 65,000 homeless individuals was within the range that was generally thought to be credible. Moreover, the standard error here implied a level of precision that was acceptable. The SPA standard errors varied substantially as a percentage of the estimated homeless count. Some margins of error were as large as plus or minus 30%. No one was happy with the larger figures, but they were a result of resource constraints and earlier design tradeoffs that on balance still seem appropriate.

TABLE 1
*Estimated tract totals and standard errors using SPA averages*

| SPA | Count-selected | Count-sampled | Count-shelter | Total | SE |
|---|---|---|---|---|---|
| 1 | 275 (8 tracts) | 370 (11 tracts) | 419 | 2,813 | 555 |
| 2 | 347 (20 tracts) | 1,034 (60 tracts) | 1,570 | 8,816 | 1,025 |
| 3 | 802 (23 tracts) | 995 (51 tracts) | 870 | 8,091 | 1,510 |
| 4 | 5,931 (86 tracts) | 99 (6 tracts) | 4,544 | 13,429 | 935 |
| 5 | 1,496 (21 tracts) | 479 (21 tracts) | 1,613 | 5,937 | 395 |
| 6 | 3,635 (40 tracts) | 1,511 (29 tracts) | 2,017 | 13,936 | 1,420 |
| 7 | 419 (24 tracts) | 438 (39 tracts) | 912 | 4,049 | 290 |
| 8 | 506 (22 tracts) | 801 (48 tracts) | 1,686 | 7,315 | 460 |
| Total | 13,411 (244 tracts) | 5,727 (265 tracts) | 13,631 | 64,386 | 4,875 |

### 8.2. Estimates at the tract level

For the purpose of implementing the second and third imputation methods, we spent considerable time examining and identifying variables commonly associated with homelessness. These included such United States census data variables as median income and the percentage of housing units vacant, and such land use variables as the percentage of land devoted to commercial activities, industry, or residences.

By and large, homelessness correlates performed as expected: homeless counts were higher, for example, in tracts with lower median income, in tracts with a higher percentage of vacant dwellings, and in tracts where a higher percentages of land was used for commercial or industrial purposes. These relationships were sometimes highly non-linear, suggesting a possible tipping effect. Thus there is a suggestion of thresholds that, if passed, tip a census tract dramatically toward having a large number of homeless people. The previous studies we reviewed assumed linear or quadratic regression relationships (e.g., Hudson [6]) and might be usefully revisited.

The modeling technique of choice, success-driven, was that of random forests (Breiman [1]). First of all, the relationships between some of the predictors and the counts were non-linear in ways that we could not anticipate and then easily capture in parametric form. Additive smoothers constructed through generalized additive models looked more promising, but abrupt changes in the counts over some small regions of the predictors led to over-smoothing in some places and under-smoothing in others. The flexibility of random forests seemed best suited to these data, and we adopted this methodology. An outline of the random forests procedure is provided in Appendix A.[4]

Several random forest models performed similarly and, for many purposes, it would not matter a lot which one was used for imputation. They differed a bit in the homelessness correlates used, but not in any surprising manner. For example, median income tends to be lower when the percentage of renters is higher; there is some redundancy between the two, and it does not matter greatly which correlate is used. The differences seen in various (sensible) models looked at are not likely to matter much for most policy purposes.

It should be emphasized that the model search was never intended to be exhaustive. We relied more on predictors thought of as familiar in this context than on the elicitation of predictors by data-driven selectors. In all, perhaps a dozen models were considered with care. To then single out a model to provide the homeless count by tract, we invoked several criteria — among them were the faithfulness of predictors when compared to observed counts at the tract level, their face validity in the homeless context in Los Angeles, the harmony of predicted totals with outside estimates of homeless numbers in Long Beach, Pasadena and Glendale, and the parsimonious nature of the model.

We ultimately used just three predictors of homeless count by tract: median household income, the percentage of dwellings that are vacant, and the percentage of land used for residential purposes. This choice of correlates is referred to as Model 1. For the sake of some comparisons, two of its competitors are brought in here.

---

[4]Analyses involving random forests were performed using the "randomForest" library in R, maintained by Andy Liaw.

| SPA | Model 0 | Model 0 SE | Model 1 | Model 2 | Model 3 |
|-----|---------|------------|---------|---------|---------|
| 1 | 2,813 | 555 | 2,707 | 2,557 | 2,361 |
| 2 | 8,816 | 1,025 | 9,220 | 9,400 | 9,768 |
| 3 | 8,091 | 1,510 | 7,505 | 7,538 | 7,603 |
| 4 | 13,429 | 935 | 16,042 | 16,378 | 16,249 |
| 5 | 5,937 | 395 | 5,634 | 6,051 | 5,720 |
| 6 | 13,936 | 1,420 | 12,453 | 12,377 | 11,898 |
| 7 | 4,049 | 290 | 5,673 | 5,003 | 5,561 |
| 8 | 7,315 | 460 | 9,175 | 9,268 | 8,586 |
| Total | 64,386 | 4,875 | 68,409 | 68,568 | 67,746 |

The first, Model 2, invokes median household income, percent residential, percent industrial, together with latitude and longitude; the second, Model 3, adds percent owner occupied, percent vacant, and percent minority to the predictors of Model 2. In all, the models incorporate three, five and eight predictors.

When the model tract predictions are aggregated to SPA and county levels, Table 2 results. Model 0 standard errors are listed alongside their predictions and provide us with a proper sense of accuracy achieved. Viable standard errors are not available for the other models, but it seems safe to allow sampling standard errors to serve as reasonable upper bounds on the models, as they bring in additional information. Recall here that Model 1 is, in fact, our choice among Models 1, 2, and 3.

When compared with Model 0 predictions, the other models give predicted totals that are well within the respective sampling errors in SPAs 1, 2, 3, 5, and 6. Differences seen in SPA 4 might be traced to the heavy reliance placed on the sample of just six tracts for predicting street count in unsampled tracts; there is some evidence to support the (higher) model answers given here.

In SPAs 7 and 8, it may be that one sees the effect of non-participation by Long Beach in the identification of selected tracts, and in the selection of sample tracts; the sample extrapolation thinks of Long Beach as a participant, while the models pick up on characteristics of Long Beach tracts and propose a higher count accordingly. In particular, the total increase of some 4,000 homeless over the sample estimate, according to each of the three models, is largely accounted for by increases in SPAs 7 and 8.

## 9. The aggregation of modeled tract counts

Tract counts gain interest as they are aggregated to recognizable levels. At the larger SPA level, Table 2 would have it that there are no incisive contrasts between the four models listed. Still, Service Provider Areas are hardly the stuff of legend, so we look to predictions of homeless populations in some incorporated cities that require less tract aggregation, yet form better known entities.

Table 3 gives estimates for eight cities chosen from SPAs 1 through 8, and presented in that order. (To produce these numbers, we have shared down model numbers according to percentage of tract area in a city — it happens occasionally that different cities share a given tract in significant proportions.) A peculiar and relevant feature of cities in Los Angeles County is that many of them are completely surrounded by the city of Los Angeles itself and, in any event, city boundaries go unmarked in large part.

TABLE 3
*Model comparisons using some estimated city totals*

| Cities | Model 0 | Model 0 SE | Model 1 | Model 2 | Model 3 |
|---|---|---|---|---|---|
| Palmdale | 662 | 134 | 517 | 558 | 434 |
| Burbank | 401 | 99 | 431 | 504 | 453 |
| Alhambra | 502 | 145 | 474 | 411 | 408 |
| West Hollywood | 128 | 26 | 147 | 166 | 156 |
| Santa Monica | 1,496 | 48 | 1,497 | 1,535 | 1,501 |
| Lynwood | 762 | 164 | 701 | 658 | 656 |
| Whittier | 403 | 29 | 415 | 428 | 412 |
| Inglewood | 971 | 67 | 1,095 | 1,154 | 1,160 |

TABLE 4
*Model comparisons using totals for non-participating cities*

| Cities | Model 0 | Model 0 SE | Model 1 | Model 2 | Model 3 |
|---|---|---|---|---|---|
| Glendale | 627 | 196 | 741 | 800 | 1,000 |
| Long Beach | 2,041 | 184 | 3,143 | 2,902 | 2,381 |
| Pasadena | 606 | 220 | 624 | 563 | 536 |

Table 4 is a continuation of Table 3, but it features the non-participating cities where predictions were sought despite the absence of the relevant tracts in the sampling universe.

Arguably, Model 1 does the best job of the three prediction models in tracking the sampling Model 0 numbers. Still, we do not feel the need to make a strong case for this on the grounds that the model is indeed competitive with the others, and simpler.

There is the suggestion that Models 2 and 3 are overly sensitive to location. Inglewood, for example, abuts the South Central Los Angeles area and has close to the highest minority population of any city in the country – only 19% White in Census 2000 – of whom a significant number were Hispanic. At the same time, it is a city with a sizable percentage of homeowners (there were some 11,000 single family owner-occupied homes in a population of roughly ten times that many people in 2000), and it holds to a fair amount of stability. Here Models 2 and 3 overshoot the sampling estimate by a considerable margin.

An overarching comment is that there can be little doubt that local homeless policy plays a large role in the ability to effect their census. Thus, Santa Monica attracts homeless residents by paying close attention to them, by taking a rather liberal stance in their behalf, and by providing a relatively safe and clean environment.

For consistency with Table 3, we have used the same tract assignment file to produce Table 4 for the three non-participating cities. A second file detailing such assignments was the one used for "official" counts, see Comment 4 in Section 11 for more details about this.

Here, Model 1 looks most like the sample estimates built on SPA averages in Glendale and Pasadena. In the case of Long Beach, use of the SPA-wide average provides a decidedly low sample estimate in that previous study of Long Beach homeless had come up with a population size of close to 6,000. A shelter count of 296 was reported for Long Beach, but this figures to be considerably short of a full census. Even more to the point, the lack of pre-identified, populous tracts in Long Beach ensured a heavy undercount. Then Model 1 does the best job of recognizing, while not correcting, the situation brought on by non-participation.

## 10. A model validation study

Setting tract-level predictions alongside the counts that were obtained on the sampled tracts, we can carry out a study of the improved accuracy of modeled totals as the aggregation sample size increases within this pool of tracts. The results can be taken as suggestive of what various consumers will encounter when they rely on modeled totals across unsampled tracts, of interest for political and other reasons. Moreover, it is instructive to carry out a validation exercise not commonly available in practical settings — one in which "truth," in the form of the observed counts, is known.

Going further, we can and do compare the performances of different models according to the same aggregation criteria. In particular, we bring in our four models as competitors. Each of them provides predictions (in the case of Model 2, the SPA-wide average of observed counts) that can be summed over the appropriate collection of tracts.

While tract sample numbers have gone into the building of the models, we do not see that this taints the present study. We looked at perhaps a dozen models, as was discussed in Section 7, and the choice among them was built on criteria that had no overt connection with model totals over collections of sampled tracts aside from paying some attention to the comparisons visible in Table 3.

Since we combine only sampled tracts, our constructs are, of necessity, artificial. Two methods of tract aggregation are considered for the purposes of this exercise: random selection, and selection with the aid of a proximity device that is described below. Other choices are possible of course, but we have not seen much in the way of qualitative differences over alternative sampling devices.

To be specific, we start with a file that is $259 \times 5$ – tracts by models plus truth (in the form of the vector of observed counts). The 259 tracts arise from the loss of one street count, see Comment 3 of Section 11, and from the elimination of all SPA 4 tracts – the 6 that were randomly sampled plus the 33 that were incorrectly entered as sampled tracts. We go through the aggregation sample sizes 4, 8, 16, 32 and 64. For each size, we first draw samples at random and without replacement from the 259 tracts, a total of 500 of them. Total homeless in the tracts drawn is regarded as small if below the median seen in the 500 draws, as large if above the median. Then, for each of the four models, we report the median absolute percent error of prediction, MAPE, as well as the percentage of cases in which predictions undershoot small totals, US, and the percentage of cases in which predictions overshoot large totals, OL. The first statistic allows comparisons to be made across models and through sample sizes, the other two quantify to some extent the imbalance inherent in the smoothing of the data through modeling.

Subsequently, for each sample size $n$, a tract number is selected at random, and a selection is made of $n$ of the $n + 8$ tracts closest in distance to it. The manner of selection here is termed geographic, and performance statistics are kept as before. The results may be seen in Table 5, all entries being percentages.

For any fixed model, Table 5 shows a rough 30% decrease in median absolute percent error on a doubling of sample size. On this scale, the model ordering is the one expected based on the size of the model, Model 0 having notably poor performance in this regard. There is a decided difference between the ability to straddle large and small targets that emphasizes smoothness more on the low than on the high end, and some differences between models on this score. When it comes to the presumably more realistic geographic constructs, models can perform quite differently at the higher levels of aggregation, though this can be an artifact of a selection method that is able to associate quite disparate tracts when distance is even moderately large. In any event, one sees in these models a tendency to overestimate the homeless when they are relatively few, to underestimate the size of the problem when there are relatively many. Referring to Table 3, an inference that the homeless problem in Whittier (15 unsampled tracts at about 10 homeless per tract) is overstated, while in Burbank (17 unsampled tracts at about 25 per tract) it is understated, is hugely tenuous, but the general point is perhaps of interest to the policy maker.

TABLE 5
*Results with aggregated census tracts*

| Size | Random selection | | | | Geographic selection | | | |
|------|------|------|------|------|------|------|------|------|
| | Model | MAPE | US | OL | Model | MAPE | US | OL |
| 4 | 0 | 46 | 4 | 32 | 0 | 39 | 10 | 29 |
| | 1 | 27 | 4 | 32 | 1 | 24 | 14 | 38 |
| | 2 | 23 | 5 | 36 | 2 | 19 | 15 | 44 |
| | 3 | 24 | 2 | 34 | 3 | 19 | 11 | 40 |
| 8 | 0 | 32 | 6 | 14 | 0 | 28 | 15 | 27 |
| | 1 | 20 | 7 | 23 | 1 | 18 | 11 | 33 |
| | 2 | 17 | 8 | 18 | 2 | 14 | 16 | 43 |
| | 3 | 18 | 6 | 18 | 3 | 16 | 10 | 33 |
| 16 | 0 | 23 | 8 | 21 | 0 | 20 | 20 | 27 |
| | 1 | 15 | 8 | 26 | 1 | 11 | 12 | 29 |
| | 2 | 13 | 8 | 28 | 2 | 10 | 18 | 37 |
| | 3 | 13 | 7 | 22 | 3 | 10 | 12 | 30 |
| 32 | 0 | 16 | 6 | 12 | 0 | 17 | 31 | 21 |
| | 1 | 12 | 14 | 17 | 1 | 9 | 13 | 26 |
| | 2 | 11 | 8 | 22 | 2 | 7 | 18 | 37 |
| | 3 | 11 | 10 | 16 | 3 | 8 | 14 | 21 |
| 64 | 0 | 11 | 10 | 10 | 0 | 8 | 41 | 23 |
| | 1 | 8 | 10 | 16 | 1 | 4 | 32 | 37 |
| | 2 | 7 | 8 | 25 | 2 | 4 | 17 | 56 |
| | 3 | 7 | 12 | 16 | 3 | 4 | 18 | 17 |

Comparing the left and right hand sides of Table 5, one sees a rather substantial increase in accuracy that attaches to the structured aggregates. Moreover, there is an increased ability of the models to pick up on high and low totals. This is comforting as one imagines that geography is not at great remove in most uses of the tract predictions.

Again in this setting, as was the case in the previous section, Model 1 holds up well against its larger competitors, and it dominates the simple Model 0 that it replaced.

## 11. Some details and caveats

1. One census tract in SPA 6 produced a street count of more than 900, and three street counts of around 600 were found in SPA 4. Thus just four tracts combined to produce more than 10% of the total observed street count, indicative of a general pattern of homeless concentrations. The pre-selection of tracts was meant to defuse the effect of such concentrations on sampling standard errors. In SPA 5 this proved to be especially effective, local knowledge of homeless locations was accurate. At the other extreme, in SPAs 1 and 2, street counts in sampled tracts were not distinguishable from those in pre-selected tracts, and no benefit was obtained from the pre-selection process in those areas.

2. In view of the prominent use of the counts found in six sampled tracts from SPA 4, it is worth remarking on what was found in them. These six were quite homogeneous in homeless count, averaging less than 20, with a largest value of 40. Put these figures alongside the counts found in the 33 tracts that were mistakenly treated as sampled tracts: counts averaged more than 30, with a largest value of 265. There is at least the suspicion that the six sample values were from the low end of the distribution of homeless counts over the tracts that were not preselected.

3. In reviewing street counts, it was never determined whether a particular tract count from SPA 2 was zero or missing, so this tract was dropped from the sampled tracts. The elimination of the 39 SPA 4 tracts' street counts then took us from the initial 299 sample tracts to the 259 used in the aggregation study.

4. As background information, we were given homeless totals for Glendale (472), Long Beach (5,845) and Pasadena (879) that had been found at various times in other studies. We took the figures provided as rough guidelines for the assessment of our projections into the three areas from which tracts were not available for sampling. In Table 3, for the sake of consistency, we used a file that assigned proportions of cities to tracts according to acreage. Earlier, based on a second file that assigned tracts to these three communities, the selection of Model 1 over Model 2 was based in part on estimates that more closely resembled the numbers at hand: in the same order, 741, 3,143 and 624 under Model 1, and 800, 2,902 and 563 under Model 2.

5. Predictions made here at the tract level take no special account of the non-homogeneous nature of the county. One approach to this might consider breaking the county into zones somewhere between the SPA and county levels, then fitting different models to such zones. From observed counts only, one tentative split incorporates a "city" zone consisting of SPAs 4 and 6, a "suburban" zone consisting of SPAs 5, 7 and 8, and a "valley" zone consisting of SPAs 1, 2 and 3. The statistical evidence for such a split comes from the size of concentrations of the homeless: the first zone has the largest by far, the second zone has significant, but moderate by comparison, concentrations; the third zone consists of fairly homogeneous tracts. It could be, though, that access to Long Beach data would suggest aligning SPA 8 with SPAs 4 and 6 rather than with SPAs 5 and 7.

6. Without input to the undercount issue, we took the position that predictions applied only to the countable population, and we harbor no illusions that all eligible persons were indeed found. The homeless population of Los Angeles is often reported to be in the neighborhood of 85,000 to 90,000. The increase in estimated homeless is attributable to a separate telephone survey that was conducted and to estimates of the homeless in jails, hospitals and the like. We were not involved in these aspects of the study.

## 12. Conclusions

A variety of methods have been employed to estimate the number of homeless individuals in particular metropolitan areas. Experience gained in the project reinforces our belief that street counts undertaken in sampled geographical areas can be used to construct useful and cost-effective estimates. Effectiveness is enhanced by sampling strategies that take serious account of available information on locations where the homeless tend to congregate. Some efficiency of estimation was lost to political pressure in the Los Angeles project, but not to any substantial degree.

A major consequence of sampling can be the need impute counts for areas not selected. For this study, estimated counts were required for each census tract because stakeholders were planning to combine census tracts in ways that could not be anticipated. We suspect that stakeholders in other areas will often want similar flexibility.

The imputation of counts to unsampled tracts can benefit from modeling in this context. We do not, however, believe that any specifics of the present choices carry over to other times or places. One can easily find, as we have in a related project, that what is "known" in advance about the location of homeless individuals and the correlates of homelessness is inaccurate. That, in turn, will affect the details of how the imputed counts are constructed. New projects are likely to profit most from up-to-date local information and a reasonably clean slate.

Aggregation of model predictions over sampled and unsampled units can improve accuracy that in this case is sufficient to help stakeholders. But a lot depends on which areas are combined, not just the number of areas. A key factor is that the fitted values tend to overestimate low counts and underestimate high counts. Bias can be a problem, but perhaps it is a mild one when an aggregate covers a decent mix of what figure to be high and low count tracts.

Finally, the level of professionalism exhibited in the field and during follow-up operations in the Los Angeles project was substantial and vital to the quality of the results. Given this, we can take the view that the estimations produced are of real value. Indeed, the counts were forwarded to HUD shortly after the data analysis was completed.

## Appendix A: The random forest algorithm for fitting conditional means

1. Take a random sample of size $n$ with replacement from the training data. The selected observations will be used to grow a regression tree. The observations not selected are saved as the "out-of-bag" (OOB) data.

2. Take a random sample of predictors.

3. Partition the data using CART as usual into two subsets minimizing the error sum of squares.

4. Repeat steps 2-3 for all subsequent partitions.

5. Compute the conditional mean for each terminal node.

6. Drop the OOB data down the tree and assign the conditional mean to OOB observations depending on the terminal nodes in which these observation land.

7. Repeat steps 1-6 a large number of times.

8. For each observation, average the OOB fitted conditional means over trees. These are the fitted values for each observation.